# The LUX Dark Matter Search – Status Update

S. Fiorucci, for the LUX Collaboration

Brown University, Dept. Of Physics, 182 Hope St, Providence RI 02912

fiorucci@brown.edu

**Abstract**. We report on the design, construction and commissioning of the Large Underground Xenon (LUX) dark matter detector at the Sanford Laboratory in Lead, SD, USA. From its inception in 2007, to its construction at a surface laboratory in lead in 2009-2010, its operation in 2011, and its re-installation 1 mile underground in 2012, we review the relevant achievements already obtained and give an outlook on how LUX will become the most sensitive detector in the field in 2013.

## 1. The LUX Detector

The large Underground Xenon (LUX) detector is a dual-phase xenon time projection chamber for the direct detection of dark matter particle interactions. With a total target mass of 370 kg, LUX is currently the largest detector of its kind. Aside from its larger size, the LUX detector possesses a number of particularities which contribute to make it extremely competitive. For more details on all technical aspects of the LUX detector the reader is encouraged to consult the recently published overview paper in Ref. [1]. Of particular note are the following:

- A double-walled cryostat made of very-low radioactivity titanium, the product of a LUX research effort [2], contains the xenon and internals.
- 122 Photo-Multiplier Tubes (PMT) with a 2-inch circular window are used in two arrays on top and bottom of the active xenon volume, in order to collect light signals. Their low radioactivity content [3] and excellent quantum efficiency allow LUX to control backgrounds and obtain outstanding light collection efficiency, measured at $> 2\times$ better than the nearest competitor –see section 3 below for details on results.
- A cryogenic system based on the liquid nitrogen thermosyphon technology, provides several hundred watt of cooling power at liquid xenon temperatures [4]. Dual phase heat exchangers lower the need for liquid nitrogen dramatically, with an efficiency measured at $> 94\%$ for a flow rate up to 350 kg/d.
- An automatic, in-line xenon sampling system allows the collaboration to verify the purity of the xenon at various points in the circulation loop, with sensitivities to levels in $N_2$, $O_2$, $CH_4$ and Kr down to 1 ppb, 160 ppt, 60 ppt and 0.5 ppt, respectively [5,6].
- A mechanism is in place for doping the xenon on demand within the circulation loop, with radioactive isotopes such as $^{83m}$Kr or $^3$H [7]. It provides a controlled source of low-energy interactions distributed throughout the entire xenon volume, and is a very effective tool to calibrate the detector.
- Shielding against ambient gammas and neutrons is provided by an 8 m diameter, 6 m tall tank filled with constantly purified and de-radonized water. The water tank is also instrumented as a Cherenkov veto, allowing the rejection of events in coincidence with rare cosmic muons traversing the tank.

## 2. Commissioning of the LUX detector

The LUX collaboration was formed in 2007 and fully funded jointly by NSF and DOE in 2008. Its history is intimately linked to that of the new Sanford Underground Research Facility (SURF) in Lead, South Dakota.

For the first three years, the lab did not exist as a research facility and the R&D effort for LUX was focused at Case Western Reserve University with the "LUX 0.1" program [8]. This allowed us to build and operate a full-size prototype of the LUX detector, with an emphasis on cryogenics, gas circulation, light readout and slow-control. It also provided a tremendous learning and training opportunity for the students and postdocs who would eventually build and operate the real detector.

By June 2010, a dedicated Surface Laboratory at SURF was completed to host the construction and commissioning of the LUX detector. This facility is as much as possible a 1:1 replica of the underground laboratory, including a class 1,000 clean room and a (smaller) water tank. It allowed the collaboration to fully test most systems while the underground facility was being excavated and outfitted. The LUX detector was assembled from the inside out between 2010 and 2011 within the clean room, while external systems, such as the xenon circulation loop, data acquisition electronics, xenon storage system, and nitrogen distribution, were built and tested just outside. In May 2011 the detector assembly was completed with 30 PMTs installed. A first cryogenic test was performed, successfully bringing the internal temperature down to 190 K and keeping it stable over a week, with the detector filled with argon gas. By October 2011 the LUX detector had been outfitted with the rest of the 122 PMTs and final upgrades to internal and external systems were completed. Between October 2011 and February 2012 LUX undertook "Run 2", during which the detector was immersed in the water shield, the full 370 kg xenon load was condensed, circulation/purification was started and data were acquired. The results were very satisfying (see section 3 below and Ref. [9]). Only a very limited number of fixes and upgrades to perform before underground re-deployment, were identified. These were corrected between March and July 2012.

In June and July 2012, the entire experiment was packed up and prepared for transport underground. The Davis Laboratory, 4850 feet underground, was completed in June 2012. The two-story laboratory space includes the 8 m diameter, 6 m tall water tank, meant to provide shielding for the LUX detector, as well as a clean room, a control room, a liquid nitrogen storage and a water filtration unit. Between July and September 2012 all of LUX was transported and reinstalled in the Davis Laboratory. The water tank was filled with water in late October, and LUX started taking Cherenkov data with the detector at vacuum. On December 10, LUX officially started "Run 3" by introducing ultra-pure xenon gas into the detector, with the goal to cool down and condense the full 370 kg by the end of January 2013.

## 3. Performance

LUX "Run 2" at the SURF Surface Laboratory achieved a number of remarkable results and demonstrated that the experiment was ready to move underground and start looking for dark matter. For more details the reader is encouraged to consult Ref. [9], but the following points are worth emphasizing:

- LUX achieved stable operation for over 100 days of running, circulating xenon at 35 slpm (equivalent to ~300 kg/day) with >98% heat exchanger efficiency, resulting in a total heat load <5 W.
- After 2 months of circulating, LUX achieved a xenon purity of 205 μs, corresponding to an electron drift length of ~25 cm, or half a detector. While not optimal, this number comes with the major caveat that a leak in the circulation path was discovered at the beginning of the run, leading to non-uniform mixing of the xenon and relatively ineffective purification.

- Light collection was measured with no applied electric field, at 662 keV ($^{137}$Cs gamma line), at 8.0 photo-electron / keV. This number is more than a factor 2 higher than that reported under similar conditions by competing experiments [10]. In terms of absolute probability for detection of scintillation photons, this corresponds to $\alpha_1 \sim 15\%$ [11]. It is very encouraging both for obtaining a low energy threshold in underground running, and for the reflectivity of PTFE in liquid xenon, which has to be >97% in order to fit the data.
- LUX measured the positively-charged muon lifetime from tagged muon interactions in the detector at $2.16 \pm 0.09$ μs, in good agreement with the world average of 2.197 μs.
- The position of individual events was reconstructed with a resolution better than 5 mm in X and Y, based on an analysis of alphas emitted from Rn daughters plated out on the gate grid wires, and from a coincidence analysis within the whole xenon volume of $^{214}$Bi – $^{214}$Po radioactive decays.

## 4. Outlook

The LUX underground science program will consist of two steps. In the first step, after ~1 month of circulating xenon to improve the purity, we will run some initial calibrations and get an idea of the background levels. We will then run in dark matter search mode for a period of ~60 days. The exact duration will be decided in a large part by the observed background levels, and tuning of the acquisition parameters. This will be a non-blind search, and we expect to see very few electronic recoil events in the energy region of interest (in a 100 kg fiducial volume, before cuts), and no nuclear recoil events by a comfortable margin. The result of this short campaign should already make LUX the most sensitive dark matter detector in the world, and will be available in 2013.

In a second step, after a new round of extensive gamma and neutron calibrations, LUX expects to start a new 300 day dark matter search campaign. This new search will be blind and will rely on early data and calibration data to define the cuts and search region. The estimated sensitivity at the end of the experiment is a WIMP-nucleus cross-section smaller than $2 \times 10^{-46}$ cm$^2$ for a WIMP mass of 40 GeV/c$^2$. More details on this estimate and what assumptions go into it can be found at the end of Ref. [9].

**Acknowledgements**
This work was partially supported by the U.S. Department of Energy (DOE) under award numbers DE-FG02-08ER41549, DE-FG02-91ER40688, DOE, DE-FG02-95ER40917, DE-FG02-91ER40674, DE-FG02-11ER41738, DE-SC0006605, DE-AC02- 05CH11231, DE-AC52-07NA27344, the U.S. National Science Foundation under award numbers PHYS-0750671, PHY-0801536, PHY-1004661, PHY-1102470, PHY-1003660, the Research Corporation grant RA0350, the Center for Ultra-low Background Experiments in the Dakotas (CUBED), and the South Dakota School of Mines and Technology (SDSMT). LIP-Coimbra acknowledges funding from Fundação para a Ciência e Tecnologia (FCT) through the project-grant CERN/FP/123610/2011. We gratefully acknowledge the logistical and technical support and the access to laboratory infrastructure provided to us by the Sanford Underground Research Facility (SURF) and its personnel at Lead, South Dakota.